# Multilayer Heater Nano cryotrons: Electrothermal Modeling of Geometry-Dependent Switching

Baibhari Priya Barua[1], Md Rahatul Islam Udoy[1], Shamiul Alam[2], Md Shafayat Hossain[3], and Ahmedullah Aziz[1]
[1]Department of Electrical Engineering and Computer Science, University of Tennessee, Knoxville, TN 37996, USA
[2]Holcombe Department of Electrical and Computer Engineering, Clemson University, Clemson, SC 29634, USA
[3]Department of Materials Science and Engineering, University of California, Los Angeles, CA 90095, USA
Email: aziz@utk.edu.

*Abstract*— **This paper presents a finite-element framework to study how device geometry affects the switching current of multilayer heater nano cryotrons (hTrons). The device comprises a 50 nm Au/5 nm Ti heater, a 100 nm $SiO_2$ spacer, a 23.6 nm NbN channel, and a Si substrate. Two-dimensional electrothermal simulations examine nine heater and channel width combinations using widths of 100, 500, and 1000 nm. Three-dimensional simulations show how heat flows through the layers and spreads across the channel near the heater-channel overlap. Heater electrical losses and heat transport are coupled, whereas the channel switching threshold is obtained separately from a temperature-dependent constitutive relation. Published global temperature measurements support the switching-law calibration, and heater-current measurements assess the resulting device response. The model measurement comparisons show a systematic reduction of switching-current density with heating, but also geometry-dependent discrepancies in the low-drive plateau and transition shape. A common extraction protocol is formulated for heater-current and heater-power requirements at a specified suppression level, while separating fitted device amplitudes from controlled geometry effects. The resulting comparisons connect thermal fields to design metrics while distinguishing calibration from validation and excluding transient-performance claims from the stationary analysis.**

## I. Introduction

Multilayer heater nano cryotrons (hTrons) control the switching threshold of a superconducting channel through heat supplied by a separate metallic heater. In a crosspoint implementation, a dielectric separates the crossed conductors, allowing thermal coupling without a direct electrical connection between heater and channel [1]. This combination of independent bias paths and thermally induced resistance is attractive for superconducting-to-semiconductor interfaces and reconfigurable cryogenic logic [2] , [3] . Related three-terminal nanocryotrons use a directly connected control terminal and have been studied for connecting single-flux-quantum circuits to conventional electronics [4], [5].

The circuit relevance of nanowire switches extends to memory: rowcolumn-addressed superconducting nanowire arrays demonstrate how device thresholds and bias sequences enter system operation [6] . For device design, however, a low heater-current requirement is not sufficient by itself. Current, dissipated power, and the spatial temperature distribution must be distinguished, because changing heater resistance can change the delivered power even at the same applied current. Earlier memory-cell and nanocryotron logic demonstrations provide complementary circuit contexts, while thermal logic and superconducting optoelectronic neurons illustrate further applications [7], [8], [9], [10].

Physical and behavioral models address different parts of this design problem. Baghdadi et al. developed an electrothermal description of the multilayer hTron, while Karam et al. extracted a SPICE model from switching-current and activation-delay measurements [11]. Nanowire circuit models also describe resistance and hotspot dynamics [12]. These approaches highlight the need to understand how device geometry and thermal transport properties influence switching-current behavior and the heater current and power required for switching.

At nanometer-scale interfaces, a bulk-material description alone does not identify the full thermal path. Thermal boundary resistance can separate the temperatures of adjacent layers [13] , and nanowire cooling studies emphasize the importance of energy transfer into the substrate [14]. A planar model further omits one spatial direction that can affect heat spreading. Consequently, agreement of a scalar switching curve does not by itself establish the accuracy of a three-dimensional temperature field or uniquely identify every interface parameter. Thin-film and silicon thermal-conductivity studies motivate temperature- and structure-dependent transport descriptions [15], [16], [17].

This work combines nine 2D geometry sweeps with available 3D reference simulations for an Au/Ti–$SiO_2$–NbN–Si hTron. It separates the FEM temperature-field solution from the empirical switching law, retains device-specific amplitudes with common shape parameters, and compares equal-suppression heater-current and heater-power requirements. The reported design matrices permit a quantitative assessment of heater versus channel width dependence. Explicit unit, power balance, and dimensionality checks define the scope of these conclusions; transverse profiles and interface-sensitivity results further characterize spatial temperature variation and interface effects.

The remainder of this paper is organized as follows. Section II reviews hTron operation, prior modeling approaches, thermal interfaces, and semiconductor electrothermal modeling. Section III describes the device and electrothermal formulation. Section IV defines calibration, verification, and design metrics. Section V presents the spatial temperature fields, transverse-temperature profiles, thermal interface sensitivity results, and geometry-dependent switching threshold trends, and discusses the limitations of the available 3D comparisons. Section VI concludes the paper.

## II. BACKGROUND AND RELATED WORK

### A. *Thermal Control and Switching-Current Suppression*

The heater supplies Joule power, which raises the channel temperature and reduces its available switching current. A fixed channel bias may therefore exceed the temperature-dependent threshold before the channel reaches its zero-current critical temperature. The applied channel bias, the extracted switching threshold, and the heater suppressing current are different quantities [1] , [11]. After the channel becomes resistive, its own Joule heating and external circuit can affect its subsequent evolution[12] ; those dynamics are outside the stationary, heater-driven extraction considered here. Self-heating hotspots in superconducting microbridges and coupled detector models illustrate the interaction of resistive heating and heat removal [18], [19]. Electrothermal feedback can also affect latching and recovery after a channel becomes resistive [20].

### B. *Physical Models, Behavioral Models, and Calibration*

Physics based finite-element models calculate the temperature distribution and heat transfer within the device, while behavioral models use simplified equations and experimentally extracted parameters to describe its switching behavior. Karam et al. use global-temperature and local-heater measurements to characterize the switching law and heater response, with a separate constriction treatment. Local defects or constrictions in the superconducting channel can reduce the measured switching current compared with an ideal, uniform channel [21]. Accordingly, fitted device amplitudes must not be confused with intrinsic geometry-dependent material properties. Reproducing data used for fitting establishes calibration agreement, not an independent predictive test. NbN deposition studies show process-dependent film and device properties, while depairing-current studies distinguish intrinsic limits from measured switching currents [22], [23].

### C. *Interfaces and Dimensionality*

Interface resistance represents an additional thermal impedance rather than a physical electrical connection [13] . Its influence depends on whether the interface lies along heater-to-channel transfer, channel cooling, or a competing substrate path. Exact numerical laws remain specific to material and interface conditions [14] . The three-dimensional simulations capture the spatial temperature distribution within the device, while cross-sectional views are extracted from the corresponding 3D solutions. Since the present study focuses on steady state electrothermal behavior, switching delay and recovery dynamics are beyond its scope. Modeling these transient effects may require additional consideration of electron–phonon energy relaxation in superconducting NbN [24].

### D. *Electrothermal Modeling in Semiconductor Devices*

Electrothermal modeling is also established in semiconductor-device analysis. Thermodynamic formulations couple carrier transport, heat generation, and conduction, while SOI and ultrathin-body studies examine self-heating and the influence of geometry and thermal coupling. At cryogenic temperatures, transistor models and nanometer bulk-CMOS measurements show that electrical behavior and self-heating remain temperature dependent. These works motivate consistent electrical-power accounting and explicit thermal-domain treatment; they provide methodological context rather than transport parameters for the NbN hTron. [25], [26], [27], [28], [29], [30].

## III. DEVICE AND ELECTROTHERMAL MODEL

### A. *Geometry and Modeling Scope*

Figure 1 illustrates the multilayer hTron structure, consisting of an Au/Ti heater, a $SiO_2$ insulating spacer, a superconducting NbN channel, and a Si substrate. The heater-channel overlap area is defined as $w_h*w_c$ while the channel's cross-sectional area is $w_c*tNbN$. Both 2D and 3D finite-element models are developed using consistent device dimensions, temperature-dependent material properties, thermal boundary conditions, and heater-power calculations to enable meaningful comparisons. The top-view layout and biasing convention are shown in Fig. 1(d). The key geometry and electrical parameters used in the stationary 2D simulations are summarized in Table I.

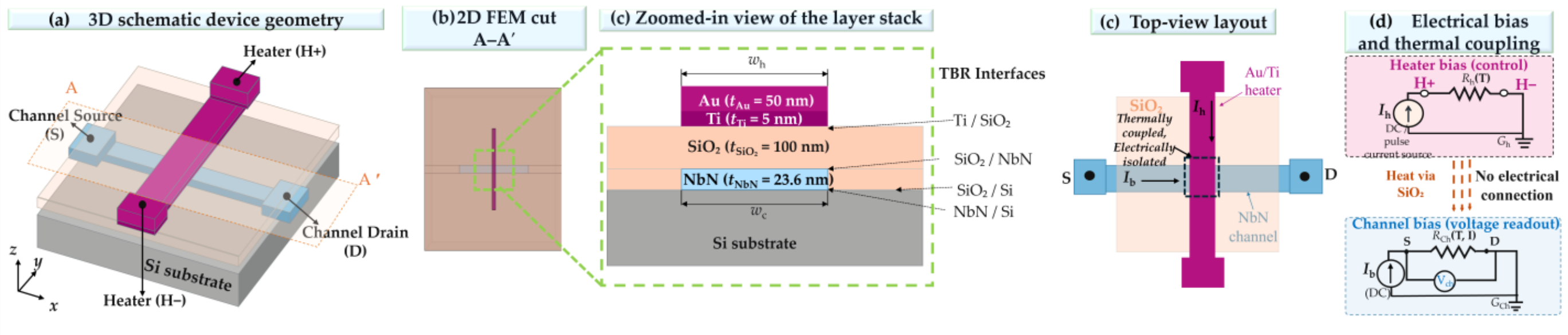

**Fig. 1: Device structure, electrical bias configuration, and thermal coupling of the multilayer hTron. (a)** Three-dimensional schematic of the crossed Au/Ti heater and superconducting NbN channel. **(b)** Two-dimensional FEM geometry showing the A–A′ section. **(c)** Enlarged cross-sectional view of the Au/Ti–$SiO_2$–NbN–Si stack, illustrating the layer thicknesses and thermal boundary resistance (TBR) interfaces. **(d)** Top-view layout and equivalent electrical bias circuits showing the independent heater and channel current paths, channel voltage readout, and heat transfer through the electrically insulating $SiO_2$ spacer. Schematics are not to scale.

**TABLE I**

Device Geometry, Material Stack, and Electrical Parameters

| Geometry and stack | | Electrical parameters | |
|---|---|---|---|
| **Parameter** | **Value** | **Parameter** | **Value** |
| $w_h$ | 100, 500, 1000 nm | $I_h$ sweep | 0–2000 μA |
| $w_c$ | 100, 500, 1000 nm | $R_{h,o}$ (reference) | 10.4 Ω |
| $L_h$ | 10 μm | αR | 0.03 $K^{-1}$ |
| tTi / tAu / $tSiO_2$ / tNbN | 5 / 50 / 100 / 23.6 nm | $\sigma Au_{,o}$ | 3.82377 × $10^7$ S $m^{-1}$ |
| Wsub × Hsub | 20 μm × 20 μm | $\sigma Ti_{,o}$ | 2.23830 × $10^6$ S $m^{-1}$ |
| $d_{oop}$ = $w_h$ (heater) | 100, 500, 1000 nm | ρNbN | 1.84 × $10^{-6}$ Ω m |

### *B. Heater Electrical Solution and Joule Losses*

The electrical behavior of the Au/Ti heater is governed by charge conservation and Ohm's law under stead-state conditions:

$$\nabla \cdot [\sigma_h(T)\nabla V_h] = 0, \qquad (1)$$

where $\sigma_h(T)$ is the temperature-dependent electrical conductivity of the heater, $V_h$ is the electric potential, and $\nabla$ denotes the nabla operator. The electric field and current density are defined as: $E_h = -\nabla V_h$, and $J_h = \sigma_h(T)E_h$ respectively. The applied heater current and electrical return boundary condition establish the current flow, while the $SiO_2$ spacer electrically isolates the heater from the superconducting channel. The heat source and total heater power are

$$Q_h = \mathbf{J}_h \cdot \mathbf{E}_h, \quad P_h = \int_{\Omega_h} Q_h \, dV \qquad (2)$$

$Q_h$ has units W/$m^3$ and $P_h$ has units W. For the same steady DC heater terminals, $P_h = I_h V_h$. The temperature dependence of the heater's electrical conductivity couples the electrical and thermal responses. As the heater temperature increases due to Joule heating, its resistance may change from its initial value at the bath temperature.

### *C. Heat Transfer and Interface Resistance*

The temperature distribution within the multilayer hTron is determined by solving the stationary heat conduction equation:

$$-\nabla \cdot [k(T)\nabla T] = Q_h, \qquad (3)$$

where $k(T)$ represents the temperature-dependent thermal conductivity of each material, T is the local temperature, and $Q_h$ is the volumetric heat generated by Joule heating in the Au/Ti heater. The heat source is applied only within the electrically driven heater region and is assumed to be zero in the remaining layers. Heat generated in the heater is transferred through the $SiO_2$ insulating layer to the superconducting channel and subsequently dissipated into the Si substrate.

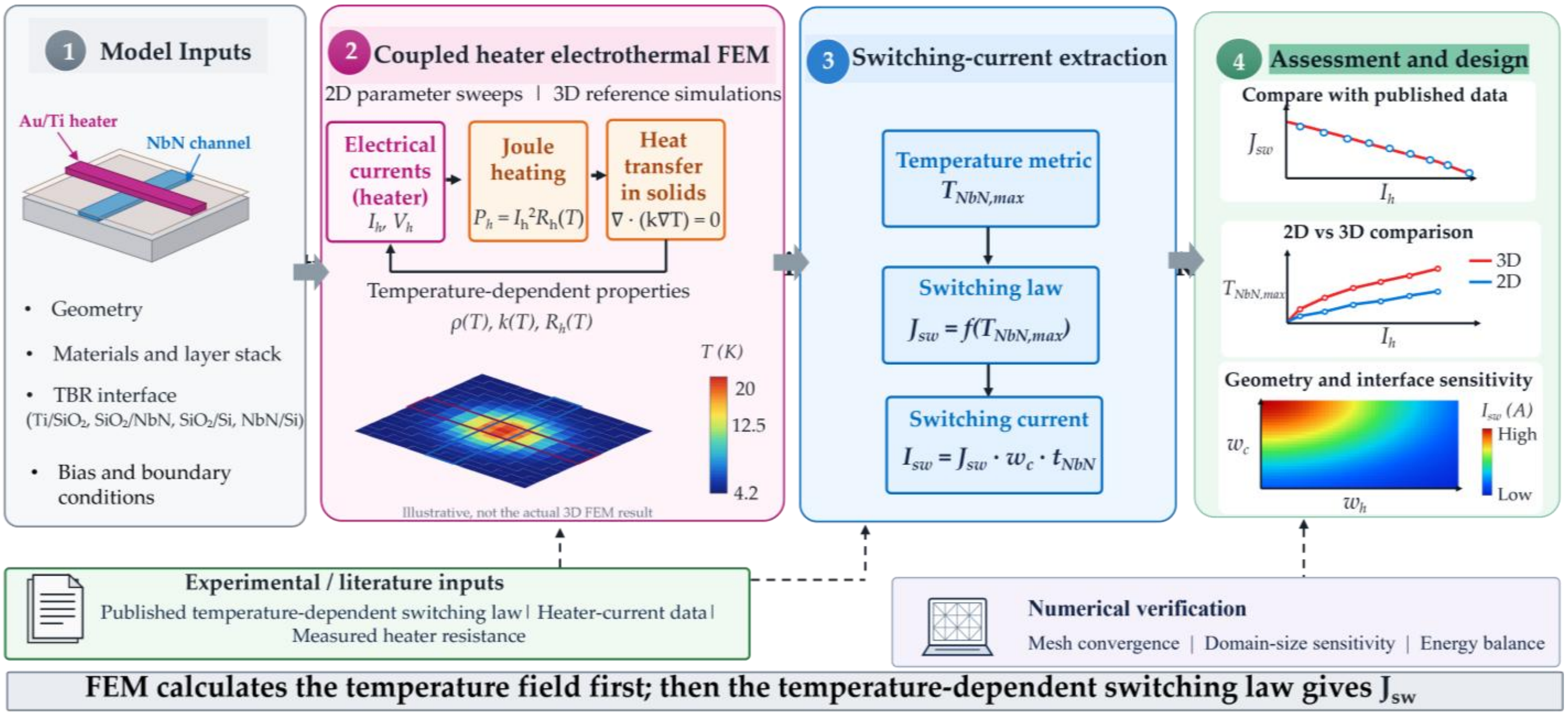

**Fig. 2: End-to-end 2D/3D electrothermal modeling and switching-current extraction workflow. (1)** Device geometry, material properties, thermal interfaces, and boundary conditions are defined. **(2)** Coupled electrical and thermal FEM simulations calculate heater Joule heating and the resulting temperature distribution, using 2D parametric sweeps and 3D reference simulations. **(3)** The maximum NbN temperature is used with a temperature-dependent constitutive relation to extract the channel switching-current density and switching current. **(4)** Model assessment includes comparisons with published measurements, 2D–3D analysis, and geometry- and interface-dependent design trends. Published experimental data supports calibration and model assessment, while mesh convergence, domain-size sensitivity, and energy balance are considered for numerical verification. The embedded plots and temperature map are illustrative.

To account for interfacial heat-transfer effects, thermal boundary resistance (TBR) is introduced between adjacent material layers. TBR allows a temperature difference across an interface while maintaining continuity of the normal heat flux. The interfacial heat flux is expressed as

$$q_{n,i} = \frac{T_{i,\text{hot}} - T_{i,\text{cold}}}{R_i''} \qquad (4)$$

where $q_{n,i}$ represents the normal heat flux across interface $i$, $T_{i,\text{hot}}$ and $T_{i,\text{cold}}$ are the temperatures on the two sides of the interface, and $R''_i$ is the area-specific thermal boundary resistance ($m^2$.K/W)[13] , [14]. The model considers the Ti/$SiO_2$, $SiO_2$/NbN, and NbN/Si interfaces to capture their influence on heat transfer and the resulting channel temperature distribution. The temperature-dependent thermal conductivities, interface coefficients, bath conditions, and reference switching-law parameters are summarized in Table II[31], [32] .

**TABLE II**

Thermal Transport, Interface Coefficients, and Switching Parameters

| **Thermal parameter** | **Value / model** | **Reference parameter** | **Value** |
|---|---|---|---|
| $L_0$ | $2.44 \times 10^{-8}$ W Ω $K^{-2}$ | $T_0$ | 4.2 K |
| kSi(T) | Interpolated; 300 at 4.2 K | Tc | 12.5 K |
| $kSiO_2(T)$ | Interpolated; 0.075 at 4.2 K | $J_0$ | $4.48 \times 10^{10}$ A $m^{-2}$ |
| kNbN(T) | max($L_0$T/ρNbN, 0.01) | p, q | 3, 2.1 |
| kh(T) | max($L_0$σiT, 0.001), i = Ti or Au | Geometry for $J_0$ | (500, 100) nm |
| **Interface coefficients and multipliers** | | | |
| Ti/$SiO_2$ | 229; 1 | $SiO_2$/NbN | 150; 1 |
| NbN/Si | 100; 0.03 | $SiO_2$/Si | 130; 1 |

*Conductivity values and floors are in W $m^{-1}$ $K^{-1}$; interpolated values are at 4.2 K. Interface entries give β (W $m^{-2}$ $K^{-4}$) and α.*

## D. *Electrical and Thermal Boundary conditions and 2D Scaling*

The electrical and thermal boundary conditions are defined to investigate both global and localized heating of the hTron. For global heating, the heater current is set to $I_h = 0$, while the bath temperature $T_0$ is varied under steady-state conditions to evaluate the temperature-dependent switching-current behavior. For localized heating, the bath temperature is maintained at $T_0$=4.2 K, and the heater current $I_h$ is swept to investigate Joule heating, heat transfer, and the resulting switching-current suppression. The electrical boundary conditions establish current flow through the Au/Ti heater, while the thermal boundary conditions account for heat transfer through the materials layers and dissipation into the substrate.

To account for the dimension omitted in the 2D model, an effective out-of-plane depth, $d_{\text{out}}$ is introduced. The total heater power is calculated by integrating the volumetric Joule heat generation over the heater cross-sectional domain and multiplying by the corresponding out-of-plane depth:

$$P_{h,\text{2D}} = d_{\text{out}} \int_{\Omega_{\text{2D}}} Q_h \, dA \qquad (5)$$

Here, $P_{h,\text{2D}}$ represents the total heater power, $Q_h$ is the volumetric Joule heat-generation rate, and $\Omega_{\text{2D}}$ denotes the electrically heated cross-sectional region. In the present model, $d_{\text{out}}$ represents the heater width ($w_h$)provided that this width corresponds to the out-of-plane dimension of the selected 2D cross-section. This scaling enables comparison of the calculated heater power across different device geometries while maintaining a distinction between the heater width ($w_h$) and superconducting channel width ($w_c$). The coupled electrothermal modeling and analysis sequence is summarized in Fig. 3.

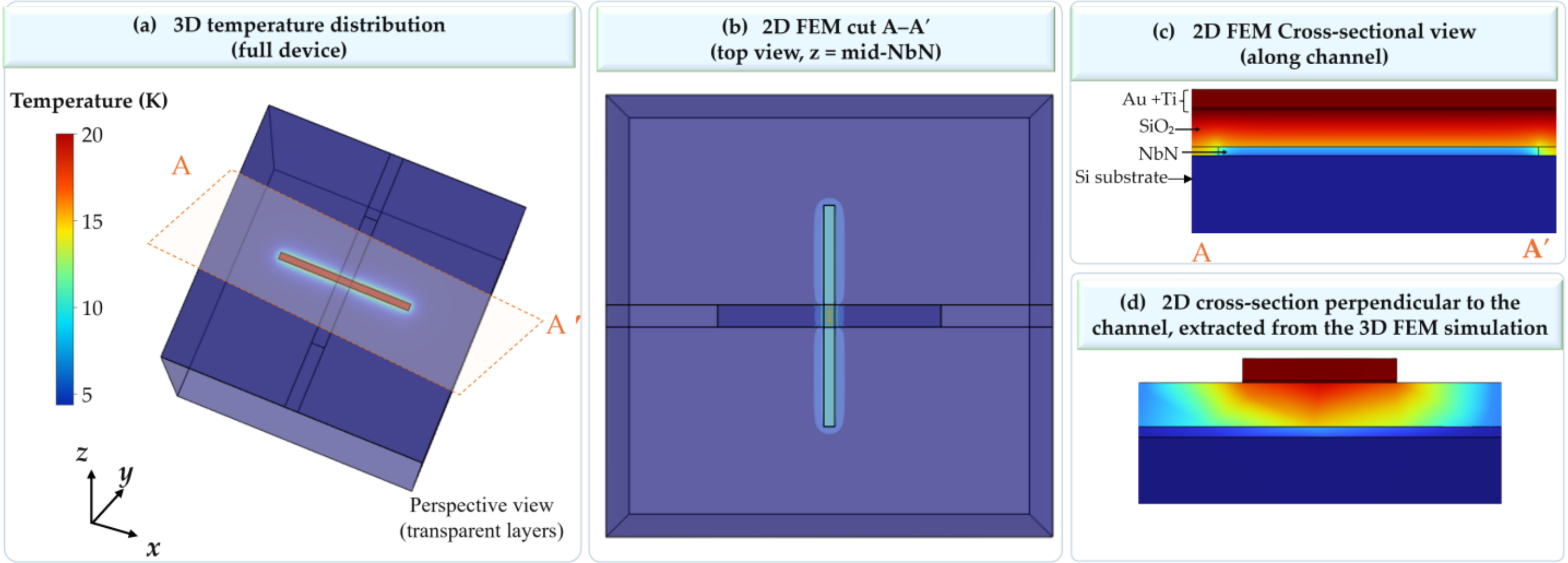

**Fig. 3: Spatial temperature distribution from 2D and 3D electrothermal FEM simulations of the multilayer hTron. (a)** Three-dimensional temperature distribution showing localized heating near the heater–channel overlap. **(b)** Top-view temperature distribution at the NbN mid-plane. **(c)** Two-dimensional cross-sectional temperature distribution along the channel (A–A′), illustrating heat transfer through the Au/Ti heater, $SiO_2$ spacer, NbN channel, and Si substrate. **(d)** Cross-sectional temperature distribution perpendicular to the channel, extracted from the 3D FEM solution, illustrating transverse heat spreading within the multilayer structure.

## IV. CALIBRATION, VERIFICATION, AND DESIGN METRICS

### *A. Switching-Current Extraction*

The switching-current density of the superconducting channel is estimated from the temperature distribution obtained through electrothermal FEM simulations. The maximum temperature within the channel is used as the effective temperature and mapped to the switching-current density using a temperature-dependent constitutive relation:

$$T_{\mathrm{eff}} = \max_{\Omega_{\mathrm{NbN}}} T, \quad J_{\mathrm{sw},g} = J_{0,g} F(T_{\mathrm{eff}}) \qquad (6)$$

Where $J_{0,g}$ represents the zero-temperature switching-current density coefficient for geometry $g$. The temperature dependence is described by,

$$F(T) = [\max\{1 - (T/T_c)^p, 0\}]^q \qquad (7)$$

Where $Tc$ is the critical temperature and $p$ and $q$ are fitting exponents. Following the reference model [11], the initial values are $Tc$=12.5 K, $p$=3, and $q$=2.1. In the present study, $J_0,g$ is determined for each of the nine geometry combinations, while a common set of p and q is used. The corresponding switching current is calculated as $I_{\mathrm{sw},g} = J_{\mathrm{sw},g}\, w_c\, t_{\mathrm{NbN}}$, where $w_c$ and $t_{\mathrm{NbN}}$ represent the channel width and thickness, respectively. Since the switching threshold is extracted from the simulated temperature field, the model does not explicitly resolve channel-current redistribution or resistive hotspot growth. Additional reductions in measured switching current caused by local defects or constrictions are considered separately [11], [21].

### *B. Hierarchical Calibration*

A hierarchical calibration approach is adopted to maintain consistency across all nine hTron geometries. Initially, the electrothermal simulations are performed using corrected heater resistances and a common set of switching-current parameters. The temperature-dependent switching relation is adopted from previously published experimental work[11], with fixed exponents $p$=3 and $q$=2.1 and a critical temperature of $Tc$=12.5 K. Subsequently, the zero-temperature switching-current-density coefficient $J_0,g$ is calibrated separately for each geometry to account for variations in the measured switching-current response. The geometry-specific coefficients are summarized in Table III, while the literature-derived exponents remain unchanged throughout the analysis.

The published global-temperature measurements provide the experimental basis for the adopted switching-current relation, whereas the heater-current-dependent measurements are used to assess the combined electrical, thermal, and switching response. This approach separates experimentally established temperature dependence from geometry-specific calibration and enables consistent comparisons across different heater and channel widths.

### *C. Suppression and Thermal-Response Metrics*

To evaluate the influence of device geometry on switching performance, the normalized switching-current density is defined as

$$s_g(I_h) = \frac{J_{\mathrm{sw},g}(I_h)}{J_{\mathrm{sw},g}(0)}, \qquad (8)$$

Where $J_{\mathrm{sw},g}(I_h)$ is the predicted switching-current density at heater current $I_h$ and $J_{\mathrm{sw},g}(0)$ is its value without heater excitation for geometry $g$. A normalized value of $s_g$=0.5 corresponds to a 50% reduction in switching-current density, while $s_g$=0.1 represents 90% suppression. The corresponding heater current and power are denoted by $I_h$,50 and $P_h$,50 respectively. These values are extracted from the simulated switching-current curves, with heater power determined from the electrical solution at the corresponding operating point.

To quantify the thermal response, the maximum temperature within the NbN region directly beneath the heater is used to define the effective overlap thermal resistance:

$$T_{\mathrm{ov,max}} = \max_{\Omega_{\mathrm{NbN,ov}}} T, \qquad R_{\theta,\mathrm{ov}} = \frac{T_{\mathrm{ov,max}} - T_0}{P_h}, P_h > 0 \qquad (9)$$

Here, $T_{\mathrm{ov,max}}$ is the maximum temperature in the heater–channel overlap region, $T_0$ is the bath or substrate temperature and $P_h$ is the total heater power. The effective thermal resistance $R_{\theta,\mathrm{ov}}$ characterizes the steady-state temperature rise per unit of dissipated heater power. Together, these metrics enable consistent comparisons of heating effectiveness and switching-current suppression across different heater and channel widths.

### *D. Numerical and Experimental Error Checks*

The agreement between simulated and experimental switching-current densities is evaluated using the normalized root-mean-square error (NRMSE):

$$\mathrm{NRMSE}_g = \frac{\sqrt{\langle e_g^2 \rangle}}{J_{\mathrm{exp,0}}} \qquad (10)$$

where $e_g = J_{\text{model,g}} - J_{\text{exp,g}}$ represents the difference between simulated and experimental switching-current densities for geometry $g$, $\langle e_g^2 \rangle$ denotes the mean squared error over the evaluated data points, and $J_{\text{exp,0}}$ is the experimental switching-current density at zero heater current.

The numerical accuracy of the thermal model is further assessed through steady-state energy conservation. The relative power-balance error is defined as

$$\varepsilon_P = \frac{|P_h - P_{\text{out}}|}{P_h}, \quad P_h > 0 \tag{11}$$

where $P_h$ is the total Joule power generated in the heater and $P_{\text{out}}$ represents the net heat flow leaving the external boundaries of the modeled domain. For zero heater power, the absolute power-balance residual is considered instead to avoid division by zero. Mesh refinement and computational-domain sensitivity are also examined to assess numerical convergence and the reliability of the 2D and 3D simulation results.

## V. RESULTS AND DISCUSSION

### *A. Spatial Temperature Distribution and 3D Analysis*

Figure 3 presents the temperature distribution obtained from the electrothermal FEM simulations. The 3D temperature maps show localized heating near the heater–channel overlap, while the cross-sectional views illustrate heat transfer from the Au/Ti heater through the $SiO_2$ insulating layer to the superconducting channel and Si substrate. These results provide insight into both vertical heat transfer and lateral temperature spreading, revealing spatial variations that cannot be captured by the maximum channel temperature alone.

The independently solved 2D model captures heat transfer within the selected cross-sectional plane, whereas the 3D model resolves temperature variations in all spatial directions. The top-view and transverse cross-sections extracted from the 3D solution further illustrate the extent and nonuniformity of heating within the superconducting channel. To evaluate the influence of model dimensionality, corresponding temperature distributions and switching parameters, including the effective channel temperature, $I_{h,50}$, and $P_{h,50}$, can be compared under identical geometry, heater power, and thermal boundary conditions. This comparison helps assess the applicability of the simplified 2D model and the importance of three-dimensional heat spreading in predicting hTron switching behavior.

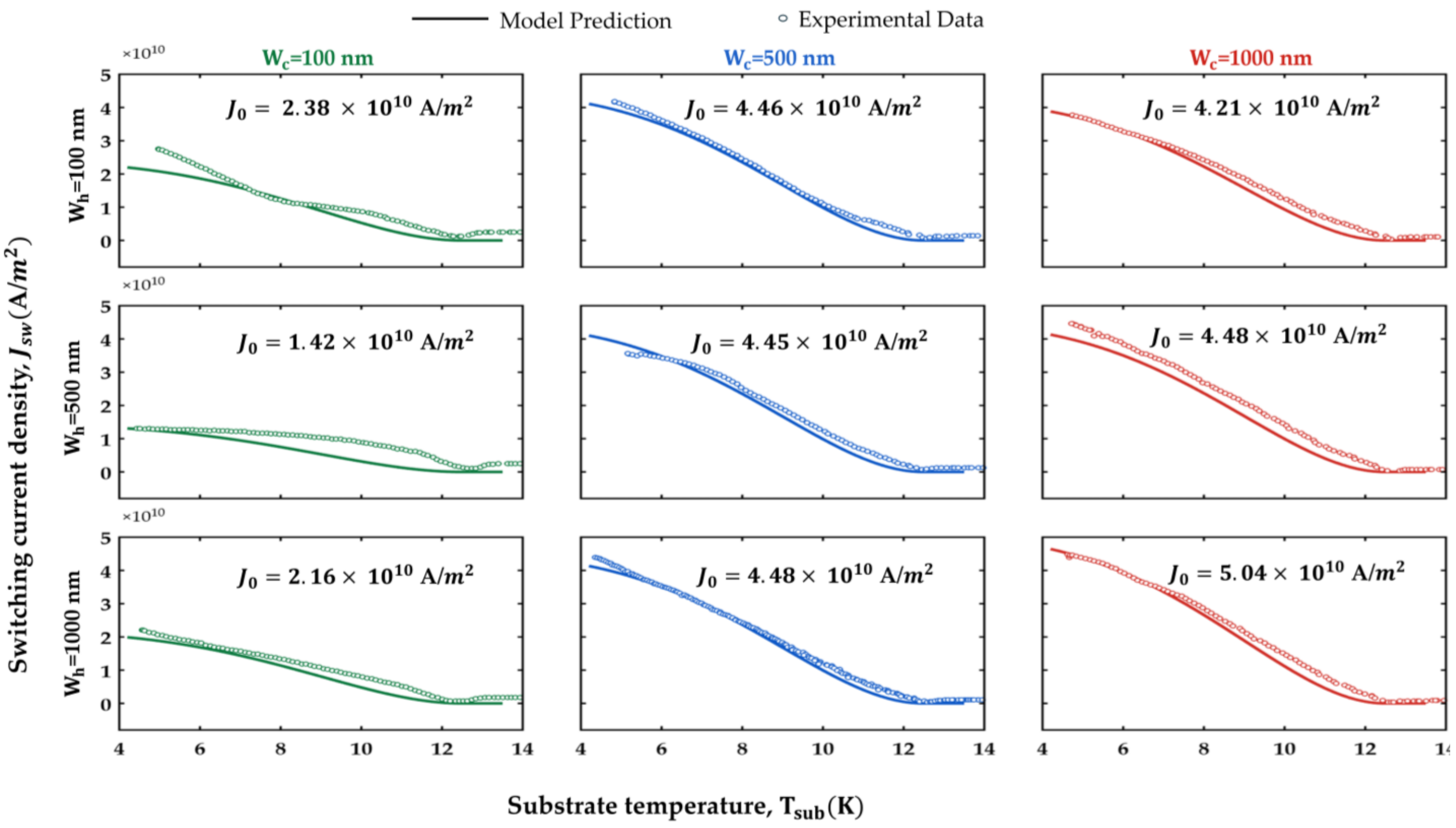

**Fig. 4: Comparison of simulated and experimental switching-current densities as a function of substrate temperature for nine hTron geometries**. Columns represent channel widths of 100, 500, and 1000 nm, while rows correspond to heater widths of 100, 500, and 1000 nm. Solid lines indicate model predictions, and open circles represent published experimental measurements from [11]. The geometry-specific zero-temperature switching-current-density coefficients $J_0$ are indicated in each panel.

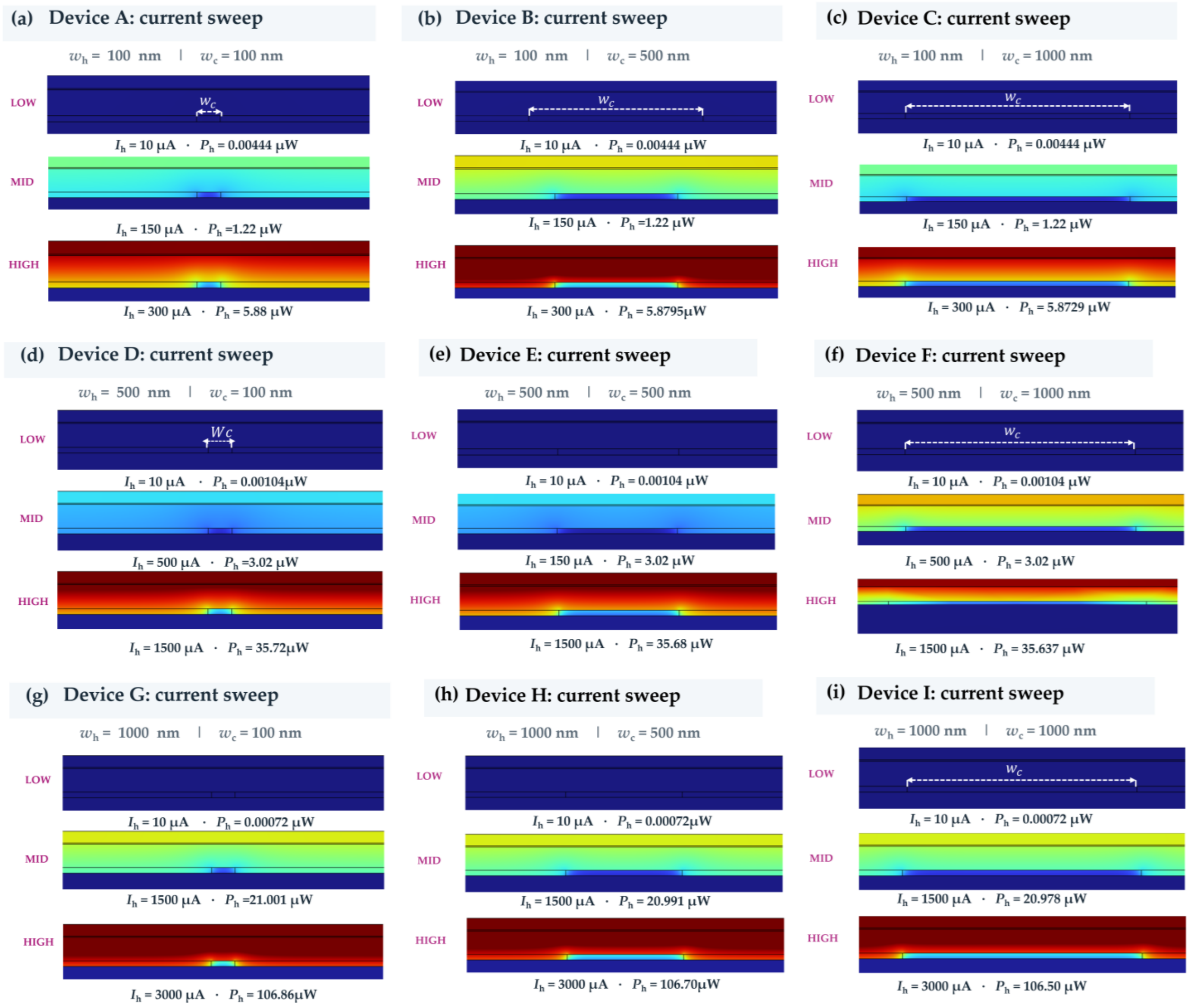

**Fig. 5: Representative cross-sectional temperature distributions obtained from 2D electrothermal FEM simulations at different heater currents. (a)** Device A with $w_h$=$w_c$=100 nm and **(b)** Device B with $w_h$ =500 nm and $w_c$ =1000) nm. (c) Device C with $w_h$ = 100 nm and $w_c$ = 1000 nm, (d) Device D with $w_h$ = 500 nm and $w_c$ = 100 nm, (e) Device E with $w_h$ = $w_c$ = 500 nm, (f) Device F with $w_h$ = 500 nm and $w_c$ = 1000 nm, (g) Device G with $w_h$ = 1000 nm and $w_c$ = 100 nm, (h) Device H with $w_h$ = 1000 nm and $w_c$ = 500 nm, and (i) Device I with $w_h$ = $w_c$ = 1000 nm. For each geometry, temperature distributions are shown at low $I_h$=10$\mu$A, intermediate $I_h$=500$\mu$A, and high $I_h$=1500$\mu$A heater currents, with the corresponding heater powers indicated. The results illustrate the increase in channel heating with applied heater current and the influence of device geometry on heat distribution near the heater–channel overlap.

### *B. Global-Temperature Switching-Current Comparison*

Figure 4 compares the simulated switching-current density with the experimental measurements reported [11] for all nine hTron geometries. The switching-current density decreases with increasing substrate temperature and approaches zero near the critical temperature of approximately 12.5 K. The wider channels exhibit relatively similar switching-current-density coefficients, whereas the 100-nm channels show lower values and greater deviations from the common temperature-dependent relation.

These differences may arise from fabrication-related variations, including local defects or constrictions in the superconducting channel, which can reduce the measured switching current [11] , [21]. Under global heating conditions, the heater current is set to zero, and the device temperature follows the applied substrate temperature. Therefore, the observed differences primarily reflect the switching characteristics of individual devices rather than heater-induced temperature variations.

The temperature-dependent switching relation adopted from [11] is used to calibrate the constitutive model. Agreement with this relation confirms its implementation, while remaining differences from the experimental measurements are assessed using the normalized root-mean-square error (NRMSE). The subsequent heater-current-dependent analysis in Fig. 6 evaluates the combined electrical and thermal response using the calibrated switching relation.

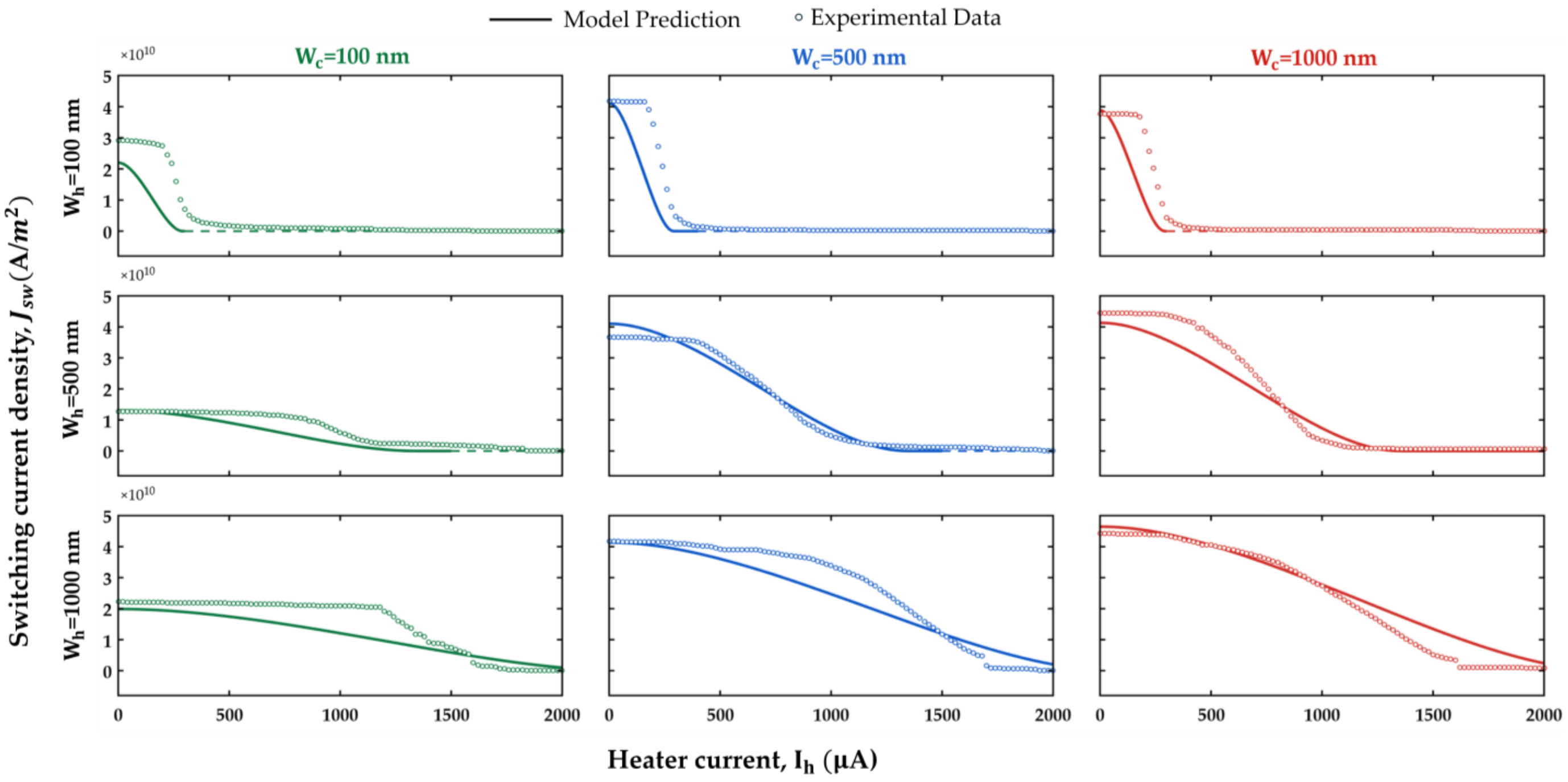

**Fig. 6**: Simulated (solid lines) and experimental (open circles) switching-current densities versus heater current for nine hTron geometries at ($T_0$=4.2) K. Columns and rows represent channel widths $w_c$ and heater widths $w_h$ of 100, 500, and 1000 nm, respectively. Experimental data are from [11].

### *C. Heater-Dependent Switching Response and Model Comparison*

Figure 6 compares the simulated switching-current density with the experimental measurements [11] for all nine heater and channel width combinations. The results show a systematic reduction in switching-current density with increasing heater current, with narrower heaters generally requiring lower currents to suppress the superconducting channel. The simulations capture this overall trend, although differences remain between the modeled and measured switching characteristics.

For devices with a 100-nm heater width, the simulated switching-current density decreases at lower heater currents than observed experimentally, and the low-current plateau is not fully reproduced. The 500-nm heater configurations show relatively closer agreement in the transition region for some channel widths, while the 1000-nm heater configurations exhibit differences in transition shape. These discrepancies may arise from variations in heater resistance, thermal transport properties, and fabrication-related imperfections. Local defects or constrictions in the superconducting channel can limit the measured switching current and produce a plateau at low heater currents.

The geometry-specific coefficients listed in Table III account for variations in the switching-current-density amplitude, while the literature-derived exponents remain fixed across all geometries. These fitted amplitudes should not be interpreted as intrinsic material-property variations caused solely by channel width. Overall, the results demonstrate that the electrothermal model captures the main geometry-dependent switching trends, while the remaining differences highlight the importance of accurate electrical calibration, interface thermal properties, and device nonuniformities.

### *D. Geometry-Dependent Thermal Response and Switching Suppression*

Figure 5 illustrates the temperature distribution at different heater currents for representative hTron geometries, highlighting the influence of heater dimensions on heat transfer to the superconducting channel. To further investigate these effects, Fig. 7 presents a comparative analysis of the thermal response and switching-current suppression across all nine geometry combinations at a bath temperature of $T_0 = 4.2$ K.

Figure 7(a) shows the maximum NbN temperature rise as a function of heater power for different heater and channel widths. The results indicate that narrower heaters produce a greater temperature rise at a given power, demonstrating more localized heating near the heater–channel overlap. In particular, the 100-nm heater requires less power to achieve a given temperature rise than the 500- and 1000-nm heaters. This comparison highlights the importance of heater geometry in determining the thermal response, independently of differences in the applied heater current. Figure 9 presents the normalized switching-current density as a function of heater power for a channel width of $w_c = 1000$ nm. As heater power increases, the channel temperature rises, resulting in a gradual reduction in switching-current density. The 100-nm heater reaches both 50% and 90% switching-current suppression at lower power than the wider heaters. These results illustrate how the geometry-dependent temperature distribution influences the predicted switching threshold.

To summarize the overall design trends, the inset in Fig. 7(b) compares the heater current and power required to achieve 50% switching-current suppression across all nine geometries. The results show a stronger dependence on heater width than channel width within the investigated design range. Together, these comparisons provide insight into the relationship between heater geometry, thermal response, and switching-current suppression, supporting the evaluation of current- and power-efficient hTron designs.

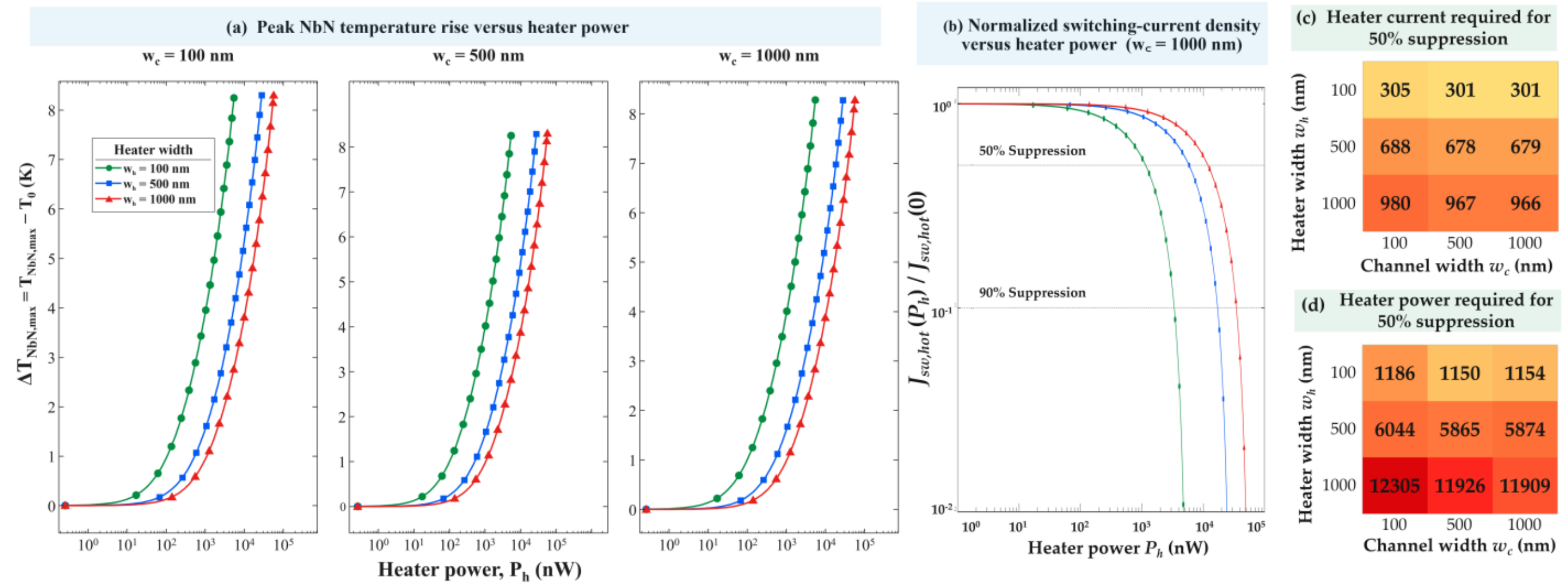

**Fig. 7**: Geometry-dependent thermal response and switching-current suppression from 2D FEM simulations at $T_0$=4.2 K. **(a)** Peak NbN temperature rise versus heater power for channel widths of 100, 500, and 1000 nm. Green circles, blue squares, and red triangles represent heater widths of 100, 500, and 1000 nm, respectively. **(b)** Normalized switching-current density versus heater power for $w_c$=1000 nm; dashed lines indicate 50% and 90% suppression. The in-set summarizes the heater current $I_{h,50}$ and power $P_{h,50}$ required for 50% suppression across all nine geometries.

## *E. Geometry-Dependent Suppression Thresholds and Electrical Consistency*

The inset in Fig. 7(b) summarizes the heater current and power required for 50% switching-current suppression across all nine geometries. Interpreting the reported current values in $\mu A$, ranges from 301 to 980 $\mu A$, while ranges from 1150 to 12305 nW. At a fixed channel width, increasing the heater width from 100 to 1000 nm increases the required current by approximately 3.21× and the required power by 10.32–10.38×. In contrast, varying the channel width at a fixed heater width produces a maximum relative spread of 1.48% in current and 3.33% in power, where the spread is defined as (maximum − minimum)/minimum. Thus, heater width has a substantially stronger influence on the reported suppression thresholds than channel width within the investigated design range.

The effective heater resistance at the 50% suppression point is calculated from the corresponding current and power as

$$R_{\mathrm{h,eff}} = \frac{P_{\mathrm{h,50}}}{I_{\mathrm{h,50}}^2} \quad (11)$$

The rounded threshold pairs yield $R_{\mathrm{h,eff}}$=12.69–12.81 Ω across all nine geometries. Because these values are derived from the same simulated current–power pairs, they provide an internal consistency check rather than an independent resistance measurement. Their near-uniformity requires reconciliation with the modeled heater lengths, terminal definitions, adopted heater-only out-of-plane scaling ($d_{\mathrm{oop}} = w_{\mathrm{h}}$), and Joule-heating integration. Until this check is resolved, the power ranking should be interpreted because of the reported 2D model rather than a verified ranking for fabricated devices.

## *F. Transverse Temperature Distribution and Interface Sensitivity*

The three-dimensional FEM simulations are used to investigate the spatial temperature distribution within the superconducting channel, particularly near the heater–channel overlap. To quantify lateral heat spreading, temperature profiles are extracted along the transverse direction at the NbN mid-thickness. The normalized temperature rise is defined as

$$\theta(y) = \frac{T(y) - T_0}{T_{\mathrm{max,line}} - T_0} \quad (12)$$

where $T(y)$ is the local temperature along the selected line, $T_0$ is the bath temperature, and $T_{\mathrm{max,line}}$ is the maximum temperature along that line. Comparing these profiles at equivalent switching-current suppression levels allows the spatial distribution of heat to be evaluated across different device geometries.

The influence of thermal boundary resistance is examined by independently varying the interfacial resistance at the Ti/$SiO_2$, $SiO_2$/NbN, and NbN/Si interfaces while maintaining consistent geometry, material properties, and operating conditions. The resulting changes in heater power required for 50% switching-current suppression, ($P_{\mathrm{h,50}}$), are used to evaluate the sensitivity of the device's thermal response. A normalized sensitivity coefficient is defined as

$$S_i = \frac{\partial \ln P_{\mathrm{h,50}}}{\partial \ln R_i''}\Big|_{\mathrm{baseline}} \quad (13)$$

where ($R_i''$) is the area-specific thermal boundary resistance of interface (i). This analysis provides a quantitative measure of how variations in individual thermal interfaces affect the heater-power requirement and helps identify the most influential thermal pathways for hTron design.

## G. Interpretation Limits

The present study evaluates steady-state electrothermal behavior and estimates switching thresholds using a temperature-dependent constitutive relation. It does not explicitly model voltage transients, resistive-hotspot growth, switching probability, or reset dynamics. A localized temperature rise therefore does not, by itself, establish these aspects of device operation. Similarly, the reported steady-state heater power should not be interpreted as switching energy, which requires consideration of the time-dependent power consumption.

Remaining discrepancies between the model and measurements, together with unresolved heater-power accounting, limit the quantitative interpretation of the results. Claims of predictive accuracy or small performance differences between geometries must be supported by documented mesh and domain convergence, assessment of uncertainty in material properties and interface laws, and quantitative 2D–3D comparisons under matched operating conditions. Accordingly, the reported results identify design trends within the assumptions of the present model rather than establish generally applicable performance predictions.

# VI. CONCLUSION

This work presents a 2D/3D electrothermal finite-element framework for studying geometry-dependent switching in multilayer hTrons. The model calculates heater-induced temperature distributions and uses a literature-based temperature-dependent relation to extract the superconducting channel's switching threshold. Across the nine geometries investigated, the 2D results indicate that the heater current and power required for 50% switching-current suppression depend more strongly on heater width than on channel width. The 3D simulations complement these comparisons by resolving the spatial distribution of heat near the heater–channel overlap.

These findings identify design trends within the assumptions of the present steady-state model. However, the nearly constant resistance inferred from the current–power pairs require verification of electrical scaling and power accounting before the reported power ranking can be generalized. Quantitative 2D–3D comparisons, transverse-temperature profiles, interface-sensitivity analysis, and numerical convergence checks are also needed to establish the accuracy and robustness of the design conclusions. The framework provides a consistent basis for connecting device geometry and thermal behavior to switching current requirements, while distinguishing calibrated threshold estimates from independently validated predictions.